\documentclass[twocolumn, aps, prb, reprint, letterpaper, notitlepage, superscriptaddress, nofootinbib]{revtex4-2}
\usepackage{amsthm}
\usepackage{mathtools}
\usepackage{physics}
\usepackage[dvipsnames]{xcolor}
\usepackage{graphicx}
\usepackage[left=23mm,right=13mm,top=35mm,columnsep=15pt]{geometry} 
\usepackage{adjustbox}
\usepackage{placeins}
\usepackage[T1]{fontenc}
\usepackage{lipsum}
\usepackage{amsmath,amssymb}
\usepackage{graphicx}
\usepackage[utf8]{inputenc}
\usepackage{csquotes}
\usepackage[colorinlistoftodos, color=green!40, prependcaption]{todonotes}
\usepackage[pdftex, pdftitle={Article}, pdfauthor={Author},colorlinks=true,linkcolor=BrickRed,citecolor=Blue,urlcolor=Blue]{hyperref} 
\usepackage{titlesec}
\usepackage{comment}
\usepackage{nameref} 
\usepackage{titlesec}
\usepackage{comment}

\begin{document}

\title{Control of an environmental spin defect beyond the coherence limit of a central spin}

\author{Alexander Ungar}
     \affiliation{Research Laboratory of Electronics, Massachusetts Institute of Technology, Cambridge, MA 02139, USA }
    \affiliation{Department of Electrical Engineering and Computer Science, Massachusetts Institute of Technology, Cambridge, MA 02139, USA }
    
\author{Paola Cappellaro}
    \email[Corresponding author. ]{pcappell@mit.edu}
    \affiliation{Department of Nuclear Science and Engineering, Massachusetts Institute of Technology, Cambridge, MA 02139, USA }
    \affiliation{Department of Physics, Massachusetts Institute of Technology, Cambridge, MA 02139, USA }
    \affiliation{Research Laboratory of Electronics, Massachusetts Institute of Technology, Cambridge, MA 02139, USA }

\author{Alexandre Cooper}
    \affiliation{Institute for Quantum Computing, University of Waterloo, Waterloo, ON N2L 3G1, Canada}

\author{Won Kyu Calvin Sun}
    \affiliation{Department of Physics, University of Illinois at Urbana-Champaign, Urbana, IL 61801, USA }
     \affiliation{Research Laboratory of Electronics, Massachusetts Institute of Technology, Cambridge, MA 02139, USA }

\date{\today} 

\begin{abstract}
Electronic spin defects in the environment of an optically-active spin can be used to increase the size and hence the performance of solid-state quantum registers, especially for applications in quantum metrology and quantum communication. Previous works on multi-qubit electronic-spin registers in the environment of a Nitrogen-Vacancy (NV) center in diamond have only included spins directly coupled to the NV. As this direct coupling is limited by the central spin coherence time, it  significantly restricts the register's maximum attainable size. To address this problem, we present a scalable approach to increase the size of electronic-spin registers. Our approach exploits a weakly-coupled probe spin together with double-resonance control sequences to mediate the transfer of spin polarization between the central NV spin and an environmental spin that is not directly coupled to it. We experimentally realize this approach to demonstrate the detection and coherent control of an unknown electronic spin outside the coherence limit of a central NV. Our work paves the way for engineering larger quantum spin registers with the potential to advance nanoscale sensing, enable correlated noise spectroscopy for error correction, and facilitate the realization of spin-chain quantum wires for quantum communication.
\end{abstract}
\maketitle

\section{INTRODUCTION} \label{sec:intro}

\begin{figure*}
\includegraphics[width = 2\columnwidth]{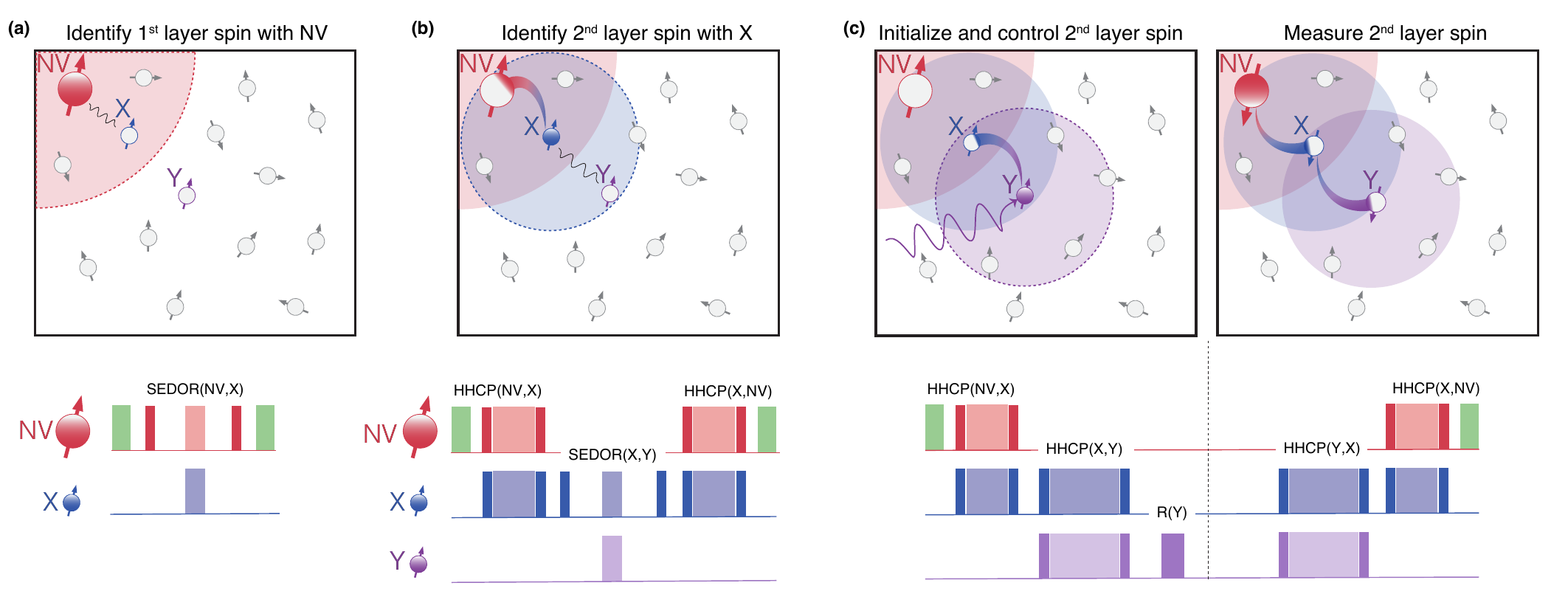}
\caption{\label{fig:1} Controlling a dark spin in the second layer to increase coherence volume and quantum register size. Our system consists of a central optically-addressable spin, the NV, and two dark electronic spins, labeled X and Y, forming a spin-chain in the configuration: NV-X-Y. 
(a)~A first-stage SEDOR sequence enables the detection of a first-layer electronic spin defect X within the coherence limit of the NV (red shading). The SEDOR sequence is comprised of a spin-echo on the probing spin with a recoupling $\pi$-pulse on the target spin. Darker (lighter) shading indicates rotations around $y$ ($x$) on the Bloch sphere. The NV is polarized and measured via laser illumination (green pulse).
(b)~A second-stage SEDOR sequence uses X as a probe to detect and characterize a second-layer spin Y, which exists outside the coherence limit of the NV. X is initialized and read out via the NV using HHCP, which consists of simultaneous spin-locking two dipolar-coupled spins at matched Rabi frequencies. 
(c)~Y is initialized by performing sequential HHCP gates across the spin-chain. Resonant microwave pulses are applied to implement unitary control operations on the second-layer spin, e.g., to sense static fields in this extended detection volume by using a $\theta$-pulse at a known Rabi frequency. The state of Y is mapped back to the state of the NV with sequential HHCP gates and is then read out through the NV fluorescence.
}
\end{figure*}

Optically-active solid-state spin defects with individual control are promising building blocks for quantum information processing~\cite{Awschalom2018}. Notably for defects in diamond, the Nitrogen-Vacancy (NV) center is a leading candidate for applications in quantum sensing~\cite{Degen2019}, while the silicon-vacancy center is a key prospect for realizing efficient quantum networks~\cite{Sipahigil2016}. Still, engineering a quantum register consisting of individually controllable environmental spins surrounding a central optically-active spin enables more powerful and interesting applications. For instance, a quantum register of nuclear spins is ideal for storing and processing quantum information given its weak coupling to the environment. Such a nuclear-spin register has thus been used to demonstrate enhanced quantum memory with record lifetimes~\cite{Abobeih2018, Bartling2022}, quantum error correction~\cite{Taminiau2014}, and quantum simulation~\cite{Randall2021}.
On the other hand, a register consisting of electronic spins, which feature stronger coupling to external fields and other spins, can enable new and complementary applications. In the areas of quantum sensing and quantum device characterization, such an electronic-spin register can be used for correlated noise spectroscopy and error characterization~\cite{Rovny2022,Wilen2021,Lupke2020,Szankowski2016}, as well as high-resolution sensing in spatial and frequency domains~\cite{Casola2018}, even surpassing the standard quantum limit in sensing~\cite{Bollinger1996,Huelga1997,Cooper2020}.

Until now, electronic-spin registers comprising an NV center electronic spin (referred to herein as NV) and optically-inactive (dark) spins in its environment have been limited in size, as they only include spins which are directly coupled to the central NV via the magnetic-dipolar interaction~\cite{Degen2021,Cooper2019,Sun2022,Rosenfeld2018, Knowles2016}. Alternatively, registers consisting of multiple NVs have been limited in size to pairs of NVs that are directly coupled~\cite{Dolde2014,Neumann2010,Jakobi2016, Lee2022}. In the case of dark electronic spin registers, the coherence volume, which is defined as the volume encompassing all of the spins that can be coherently controlled, has been limited by the coherence time of the central NV. We refer to the spins within this limit as first-layer spins. To pertain to this first layer, the dipolar coupling strength between the first-layer spins and the central NV, denoted as $d(\vec r)$, must satisfy the condition $d(\vec r) \gtrsim 1/T_2$, where $T_2$ is the NV’s coherence time. If spins beyond this first layer could be accessed, the coherence volume of electronic-spin registers could be scaled up beyond the NV coherence limit. A promising approach to surpass the coherence limit is utilizing a coherent first-layer spin to detect and subsequently control a second-layer spin not directly coupled to the central spin.

Although control of nuclear spins beyond this first-layer limit has been achieved in other works~\cite{Jiang2009,vandeStolpe2023}, achieving the same control for electronic spins is a more significant challenge limiting the scalability of such registers. The electron-electron coupling strength, several orders of magnitude larger compared to nuclear-nuclear coupling, leads to more complex interaction graphs, requiring more complicated control schemes over a given coherence volume. To overcome this problem, one must decrease the spin density and expand the control to higher layers of spins. Thus, achieving control of an electronic spin beyond the first layer is a key first step towards building a practical electronic-spin register.

In this paper, we extend the coherence volume of a network of dark electronic spins beyond the coherence limit of the optically-active central NV. Following the control protocol illustrated in \hyperref[fig:1]{Fig.~1}, our approach exploits a mediator first-layer spin (X) to detect and characterize a second-layer spin (Y) using spin-echo double resonance (SEDOR) measurements~\cite{deLange2012}. We find that the central NV, X, and Y spins form a chain where Y is located outside the NV's coherence limit. Furthermore, we demonstrate how universal single-qubit control can be extended to this second-layer spin. We initialize and readout the state of Y by performing polarization transfer across the spin-chain using Hartmann-Hahn Cross Polarization (HHCP)~\cite{Hartmann1962}. We then successfully implement coherent control of Y by driving and measuring Rabi oscillations. 


{By controlling an electronic spin defect in diamond that cannot be directly sensed by the central NV, we realize the first step to controlling larger electronic-spin registers. Our approach can be scaled to detect environmental spins in the $N^{\text{th}}$ higher layers by recursively applying the HHCP$^N$-SEDOR-HHCP$^N$ control sequence shown in \hyperref[fig:1]{Fig.~1(b)}. We show that our method of concatenated polarization transfer to sense spins in higher layers is more robust to decoherence as compared to correlated spectroscopy protocols, and in fact scales exponentially better with the number of layers. By addressing spin-chains across several layers rather than a group of spins within the same layer, our method can be used to achieve selective control of a greater number of spectrally overlapping spins. This key consequence of our method together with the capability to detect new distinct defects within the same layer~\cite{Cooper2020} will offer a more scalable path for solid state quantum registers. Accessing such larger spin registers would enable preparing correlated states of electronic spins to improve metrological performance in magnetic sensing applications at room temperature, ultimately achieving a sensitivity scaling with $\sqrt{n}$ for an $n$-spin register~\cite{Goldstein2011, Cooper2019}. And by controlling longer spin-chains to reach the diamond surface, it may be possible to achieve further insights into surface spin dynamics~\cite{Dwyer2022}, and improved imaging of spin-labeled molecules~\cite{Schlipf2017}.


\section{SYSTEM HARACTERIZATION WITH SPIN-ECHO DOUBLE RESONANCE} \label{sec:system}

\begin{figure*}
\includegraphics[width = 2\columnwidth]{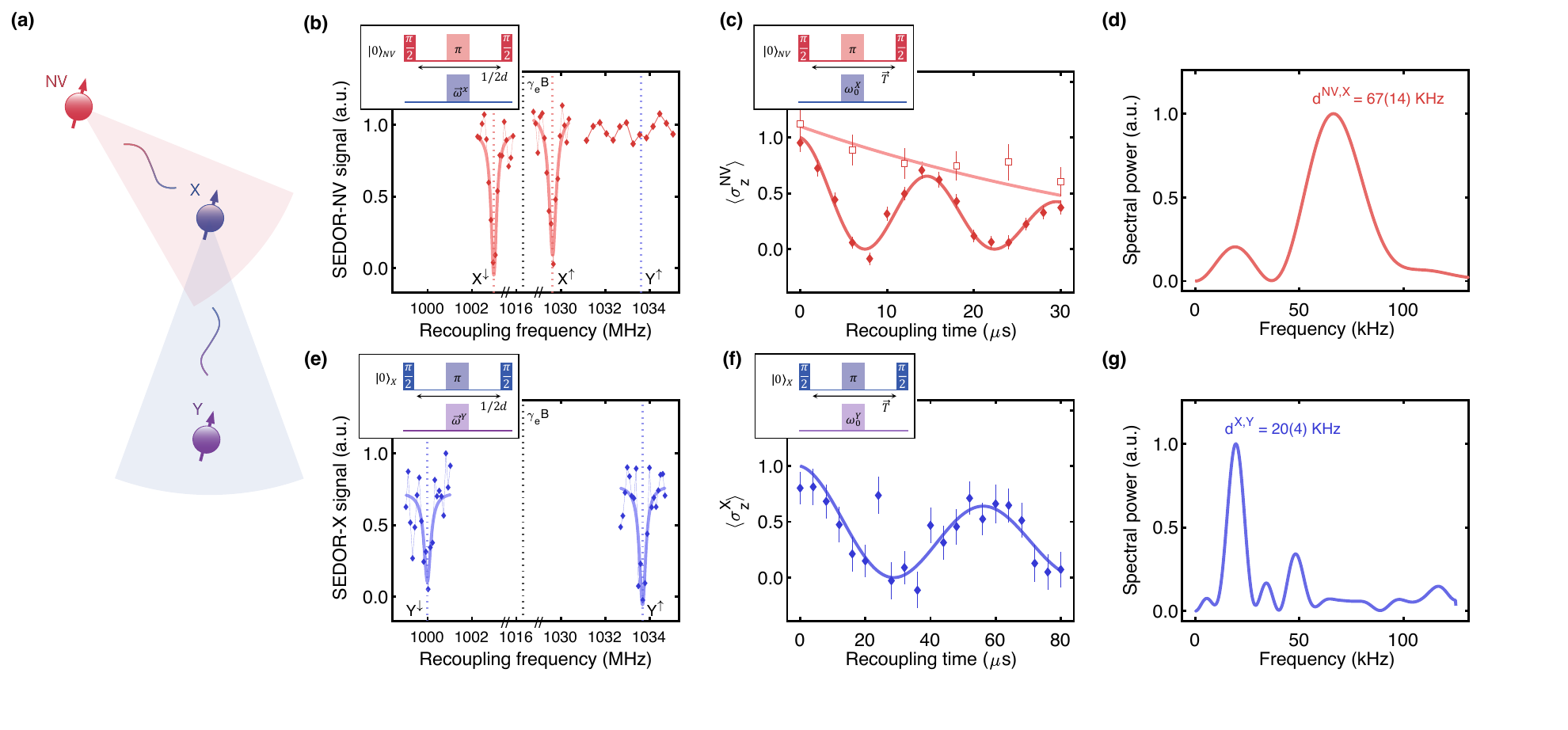}
\caption{\label{fig:2} Identification of $\omega_0^i$ and $d^{ij}$ for the NV-X-Y spin-chain using SEDOR. 
(a)~Illustration of the three-spin system, where the shaded region corresponds to the coherence limit of the probing spin (red for the NV, and blue for X). 
(b)~SEDOR-ESR on the NV reveals the resonance frequencies and hyperfine splitting of the first-layer spin X, with $A = 26.5(3)$ MHz. No resonance is detected at the Y frequency (see (e)) confirming no coupling between the NV and Y. The y-axis is $\langle \sigma_{z}^{\mathrm{NV}}\rangle$ with an added baseline correction to normalize for decoherence during the recoupling time of the sequence. 
(c)~SEDOR-Ramsey(NV,X) plotted as diamonds shows coherent oscillations mediated by the dipolar coupling between the NV and X spins. Spin-echo on the NV plotted as squares indicates the coherence time for the NV, with $T_2^{\text{NV}} = 50(10)$ $\mu s$. 
(d)~FFT of the SEDOR-Ramsey(NV,X) signal features a single dominant peak at the dipolar coupling strength of $d = $ 67(14) kHz, confirming X is a non-degenerate spin. 
(e)~SEDOR-ESR on X reveals the resonance frequencies and hyperfine splitting of the second-layer spin Y, with $A = 33.5(3)$ MHz. 
(f),(g)~SEDOR-Ramsey(X,Y) shows coherent oscillations between the two coupled spins and the FFT confirms Y is a non-degenerate spin with coupling strength $d = 20(4)$ kHz. Data in (e) and (f) are renormalized from $\langle \sigma_z^\text{NV}\rangle$ according to the calibration measurement in~\hyperref[sec:appendixC]{App.~C}. 
(b),(e)~insets: SEDOR-ESR pulse sequence comprised of a spin-echo on the probing spin (NV,X) at fixed evolution time, and a recoupling $\pi$-pulse with swept frequency. (c),(f)~insets: SEDOR-Ramsey pulse sequence comprised of a spin-echo on the probing spin (NV,X) at swept recoupling time, and a recoupling $\pi$-pulse on resonance with the target spin (X,Y). See \hyperref[sec:appendixB]{App.~B} for the empirical fitting functions and best-fit parameters.
}
\end{figure*}

Our first goal is to characterize the spin Hamiltonian for a system of interacting electronic spins in the environment of the single NV featured in our previous works~\cite{Cooper2019,Cooper2020,Sun2022}. In~\cite{Cooper2020} we characterize two electron-nuclear spin defects, labeled X$_1$ and X$_2$, which are directly coupled to this NV center (in the remainder of this work we refer to the X$_1$ spin as X). However, for completeness, we assume no prior knowledge of these spins. The experiments are performed on the same experimental setup at room temperature with an external magnetic field of magnitude $363.0(1)~\text{G}$ aligned along the molecular axis of the NV center,~$z^{\mathrm{NV}}$. Assuming non-degenerate spins, the electronic spin Hamiltonian for the network takes the following form under the secular approximation\footnote{The secular approximation is valid when $|\omega_{0}|\!=\!|\gamma_{e}B_0|\!\gg \!|\omega_d|$}: 
\begin{equation}
    \hat{\mathcal{H}} =  \frac{1}{2} \sum_{i} \omega_{0}^{i}\hat{\sigma}_{z}^{i} +  \frac{1}{4}\sum_{i \neq j,j} \omega_d^{ij} \hat{\sigma}_{z}^{i} \hat{\sigma}_{z}^{j},
\end{equation} 
where $z = z^{\mathrm{NV}}$ defines the orientation of the external magnetic field, $\omega_{0}^i = \gamma_eB_0 + m_{I}A^i$ corresponds to the electron-Zeeman splitting and hyperfine shift in the $m_{I}$ nuclear spin manifold for the $i$-th spin, and $\omega_d^{ij} = 2\pi d^{ij}$ corresponds to the dipolar coupling strength between the $i$-th and $j$-th spins. We begin the characterization protocol detailed in~\hyperref[fig:1]{Fig.~1(a),(b)} by using the central NV as a probe to detect and identify the potential coupling to a first-layer spin (X) via spin-echo double resonance (SEDOR). Once a first-layer spin is characterized by measuring $\omega_0$ and $d$, this spin can then be used as a probe to detect a second-layer spin (Y) or its coupling to other first-layer spins. This protocol can be extended to higher layers in order to construct the interaction graph for the spin network, i.e. measure all terms in the secular spin Hamiltonian. 

The SEDOR sequence requires a mechanism for initialization and readout of the probing spin (see \hyperref[sec:appendixA]{App.~A}), which for the NV can be accomplished by applying a laser pulse. However, the environmental dark spins do not feature spin-dependent optical transitions. Thus in order to harness them as probing spins, polarization must be transferred from and to the NV for initialization and readout. This is accomplished by using the Hartmann-Hahn Cross Polarization (HHCP) sequence, which consists of simultaneous spin-locking both spins at matched Rabi frequencies~\cite{Hartmann1962}. Polarization can be transferred to a dark spin beyond the first layer by concatenating a sequence of HHCP blocks starting with the NV. HHCP between the NV and the previously identified first-layer spins is characterized in detail in our previous works~\cite{Cooper2019, Cooper2020}. While correlation spectroscopy could be used to detect spins in higher layers without polarizing the probing dark spin~\cite{Schweiger2001}, we ultimately intend to harness these spins in a quantum register, requiring their spin state to be initialized and measured. 

\subsection{Characterization of a first-layer spin}\label{sec:systemA}

Starting with the NV as a probe, we seek to detect a first-layer spin by performing SEDOR(NV,X). The pulse sequence is outlined in \hyperref[fig:1]{Fig.~1(a)}, and consists of a decoupling spin-echo on the NV and a recoupling $\pi$-pulse on the potential X target spin. For SEDOR-ESR(NV,X), the frequency of the $\pi$-pulse is swept around the free electron Larmor frequency, $\gamma_e B_0$, to identify a resonance corresponding to an electronic-spin transition. The signal $\langle \sigma_z^{\mathrm{NV}} \rangle$ is extracted from the NV fluorescence by performing a differential measurement and normalizing by the NV fluorescence contrast from the $m_s = 0$ and $m_s = -1$ states (see our earlier work~\cite{Sun2022} for details of this procedure as well as the error analysis). The SEDOR-ESR signal decreases around the resonance frequency and indicates the existence of a dipolar-coupled first-layer spin. As derived in~\hyperref[sec:appendixA]{App.~A}, the analytical signal as a function of recoupling frequency (target spin detuning) follows a sinc-like profile with maximum contrast when the recoupling time $T = 1/(2d^{\text{NV,X}}).$

Once a first-layer spin's resonance frequency, $\omega_0^\text{X}$, is detected in the SEDOR-ESR(NV,X) spectrum, we measure the coupling strength, $d^{\text{NV,X}}$, by performing SEDOR-Ramsey(NV,X). This consists of resonantly driving both spins and sweeping the interaction (recoupling) time. As sown in~\hyperref[sec:appendixA]{App.~A}, the analytical signal is $\langle \sigma_z^\text{NV} \rangle= \cos(2 \pi d^{\text{NV,X}}T)$, and the coupling strength is extracted by taking the FFT (see \hyperref[sec:appendixB]{App.~B} for further details).  While a resonance signal in the SEDOR-ESR spectrum at $\omega_0^\text{X}$ can result from multiple degenerate spins, a single dominant peak in the SEDOR-Ramsey FFT spectrum verifies the presence of a single spin coupled to the NV at this frequency. Other peaks in the FFT would correspond to degenerate spins at different NV-X coupling strengths.

\hyperref[fig:2]{Figures~2(b)-(d)} demonstrate this characterization protocol for a first-layer spin. In \hyperref[fig:2]{Fig.~2(b)}, we apply SEDOR-ESR on the NV, and successfully measure the electron-spin transitions corresponding to a spin X which is coupled to the NV, and features a hyperfine splitting $A = 26.5(3)~\text{MHz}$. The spectrum contains two peaks of equal intensity corresponding to the two hyperfine states, X$^{\uparrow}$ and X$^{\downarrow}$, each with half of the total NV contrast. These observations confirm that X is an electron-nuclear spin defect with S=1/2 and I=1/2 ~\cite{deLange2012, Cooper2020}. Next, we perform SEDOR-Ramsey on the NV with the recoupling $\pi$-pulse on resonance with a single X hyperfine transition. In \hyperref[fig:2]{Fig.~2(c)} we observe that this signal oscillates at the coupling strength frequency of 67(14) kHz. Further, the single dominant peak at this frequency in the accompanying FFT spectrum in \hyperref[fig:2]{Fig.~2(d)} conclusively identifies the presence of a single spin at the X resonance (within the uncertainty of the spectral width). We also determine the coherence time of the NV by performing a spin-echo decay measurement, finding a mean-decay time of $T_2^{\text{NV}} \approx 50 \; \mu$s. This coherence time sets a lower bound on the coupling strength for a first-layer spin, which is evaluated to be approximately 10 kHz. By measuring $\omega_0^\text{X}$ and $d^{\mathrm{NV,X}}$, we complete the characterization of the first-layer spin X. We now seek to exploit this first-layer spin as a probe to detect and characterize a second-layer spin.

\subsection{Characterization of a second-layer spin}

In our earlier work~\cite{Sun2022}, the presence of coherently interacting spins around X was revealed from the characterization of its decoherence, motivating further investigation to realize a larger quantum register. Now, following the above protocol, we identify a second-layer spin Y coupled to X. 

More concretely, we perform SEDOR as above with X replacing the NV as the probing spin, and refer to this sequence as SEDOR(X,Y). To initialize and measure X, we apply a HHCP gate between the central NV and X. This three-block control sequence is shown in \hyperref[fig:1]{Fig.~1(b)}. Initialization and measurement of X, labeled as HHCP(NV,X) and HHCP(X,NV) respectively, is implemented by setting the spin-lock duration to $1/(2d^{\text{NV,X}})$ which achieves an $i$SWAP gate (a SWAP gate with an additional $\pi/2$ phase)~\cite{Cooper2019}. For initialization, the laser pulse precedes the $i$SWAP gate to prepare the NV in a polarized state. For measurement, the laser pulse follows the  $i$SWAP gate for optical readout of the NV. We convert the NV polarization signal $\langle \sigma_z^\text{NV} \rangle$ to $\langle \sigma_z^\text{X} \rangle$ for the SEDOR(X,Y) results by normalizing according to the calibration measurement included in~\hyperref[sec:appendixC]{App.~C}. We drive both hyperfine transitions of X during the HHCP and SEDOR blocks. This increases the signal contrast in order to compensate for an approximately $50\%$ reduction in signal arising from control imperfections.  

\hyperref[fig:2]{Figures~2(e)-(g)} demonstrate this second-layer spin characterization protocol by identifying the coupling between the X and Y spins. The SEDOR-ESR(X,Y) signal featured in \hyperref[fig:2]{Fig.~2(e)} reveals the resonance frequencies of the two electron-spin transitions of Y, with a hyperfine splitting of $A = 33.5(3)~\text{MHz}$. Following a similar argument from the identification of X above, we conclude that Y is also an electronic and nuclear spin-1/2 defect. The SEDOR-Ramsey(X,Y) signal and its FFT in \hyperref[fig:2]{Fig.~2(f),(g)} characterize the coupling strength between the two spins, finding $d^{\text{X,Y}} = 20(4)$ kHz, and verify the presence of a single spin at the Y resonance. We remark that the hyperfine components of X as reported in~\cite{Cooper2020} are not consistent with any of the various nitrogen or hydrogen defects that have been previously reported~\cite{Zhou1996,Etmimi2009,Peaker2018}.  Further, the measured hyperfine splitting of Y at this particular external field orientation is not consistent with the possible values for the hyperfine splitting of X at any external field orientation. These observations support our claim that they are different defects (see~\hyperref[sec:appendixC]{App.~D} for further details). For complete defect identification of Y, further experiments must be performed to fully characterize the defect's hyperfine structure, as done for the X defect in~\cite{Cooper2020}, as well as to identify the nuclear spin species using electron nuclear double resonance (ENDOR) techniques~\cite{Schweiger2001}.

We experimentally verify that Y is a second-layer spin by checking that the NV is not coupled to Y within its coherence time, i.e., up to $T_{2}^{\mathrm{NV}}$ such that $d^{\text{NV,Y}} < d_{\text{min}} \approx 10 \; \text{kHz}$. A spin below this coupling strength threshold would not be detected by the NV due to the decay in contrast, resulting in a SNR $<1$ over a stable averaging duration. We therefore check whether the NV is coupled to Y above this threshold by performing SEDOR-ESR on the NV with a recoupling time of $T_2^{\mathrm{NV}}$. The result is shown in \hyperref[fig:2]{Fig.~2(b)}, and we observe no change in the NV signal at the Y resonance, concluding Y is indeed a second-layer spin. It should be noted that while dynamical decoupling can be performed on the NV and dark spins to increase $T_2$ and hence the direct coupling to more distant target spins (with weaker coupling), such as in~\cite{Taminiau2012}, detection through an intermediary spin with stronger coupling to both the NV and the target is more experimentally desirable. This is because the total sequence length for direct detection with a single two-qubit gate will be longer than concatenating the sequence into two separate two-qubit gates, as the two-qubit gate time, $T$, scales with the separation distance between spins as $r^3$.

With the Y spin fully characterized by measuring $\omega_0^\text{Y}$ and $d^{\mathrm{X,Y}}$, we complete the identification protocol of this three-spin system consisting of the chain NV-X-Y. We are now equipped to perform initialization, control, and readout of a spin outside the NV coherence limit. The detection of a second-layer spin via X is significant as it requires probing the environment beyond what the NV center can directly access, showing that we are no longer limited by its coherence time. We choose to conclude our system identification up to a second-layer spin due to setup instabilities that occur over the longer averaging duration required to probe higher layers. However with improved setup stability and control electronics, this protocol can be extended to probe for spins beyond the second layer. 

\section{CONTROL OF A SECOND-LAYER SPIN} \label{sec:ycontrol}

\begin{figure*}
\includegraphics[width = 1.8\columnwidth]{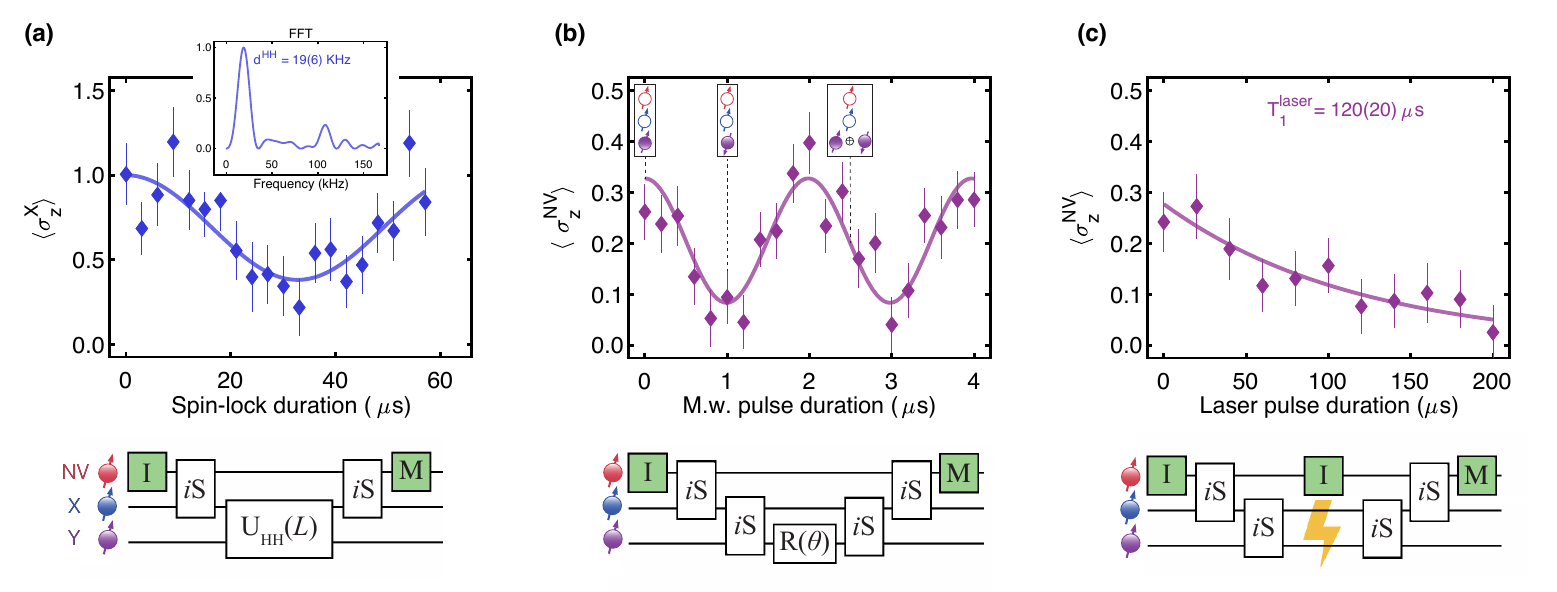}
\caption{\label{fig:3} 
Demonstrating universal control of the second-layer spin Y. 
(a)~Creating polarization transfer between X and Y via HHCP. The polarization signal $\langle \sigma_z^\text{X} \rangle$ reaches a minimum near the $i$SWAP time of $1/(2d^\text{X,Y})$ to initialize Y via cascaded polarization transfer. The signal is normalized by the method in~\hyperref[sec:appendixC]{App.~C}, with an additional offset such that the maximum fit value equals 1 (see \hyperref[sec:appendixB]{App.~B} for further details).  Inset: FFT showing the frequency of polarization transfer is at the expected dipolar coupling strength, $d^\text{X,Y}$.
(b)~Achieving unitary control of Y by driving Rabi oscillations. Y is initialized and read out by applying sequential $i$SWAP gates along the chain, with the HHCP(X,Y) gate calibrated from the signal in (a). We apply a microwave (m.w.)~pulse of swept duration on resonance with the Y electronic spin transition. The y-axis scale is the NV contrast, and creates a mapping of the Y spin state for readout. Illustrations within the plot show the polarization of the spin-chain, indicated by their shading and orientation. 
(c)~Measuring the depolarization time under optical illumination for Y, with $T_{1}^{\mathrm{laser}} = 120(20)$ $\mu$s. Over this timescale, Y can act as a memory qubit to enable repetitive readout. 
(a)-(c)~Quantum circuits below plots: the green I and M blocks represent laser pulses to initialize and measure the NV, the $i$S blocks represent $i$SWAP gates between 2 spins, the R($\theta$) block represents a pulse of length $L = \theta/\Omega_{0}^\text{Y}$, where $\Omega_{0}^\text{Y} = 0.5$ MHz .
}
\end{figure*}

As a way to scale our register to include qubits beyond the first layer of spins, we demonstrate the three steps required for universal control on the second-layer spin Y: initialization, unitary control, and readout~\cite{DiVincenzo2020}. We first characterize the initialization protocol by measuring the coherent transfer of polarization from X to Y using HHCP. This three-step sequence, as outlined in \hyperref[fig:3]{Fig.~3(a)}, consists of initializing X via HHCP(NV,X), transferring polarization from X to Y via a variable-duration HHCP(X,Y) block, and reading out X via HHCP(X,NV). We show the resulting polarization signal $\langle \sigma_z^\text{X} \rangle$ in \hyperref[fig:3]{Fig.~3(a)}, which oscillates at the expected frequency of the dipolar coupling strength $d^\text{X,Y}$~\cite{Hartmann1962}. This oscillation indicates polarization is transferred from X to either a single environmental spin or to possibly multiple spins at the Y resonance. However, we confirm this indeed polarizes the second-layer spin Y detected from the SEDOR measurements by observing that the fitted signal contrast of 0.7(1) is consistent with the SEDOR-ESR(X,Y) contrast of 0.760(6) in \hyperref[fig:2]{Fig.~2(e)} (where both signals are normalized by the same method presented in~\hyperref[sec:appendixC]{App.~C}). Thus, at the HHCP(X,Y) signal minimum, Y is maximally polarized to within our control limitations, and we achieve sequential polarization transfer across the spin-chain. We can then construct an $i$SWAP gate between X and Y to implement the initialization and readout steps necessary to achieve universal control of this second-layer spin. 

We satisfy the requirements of unitary control and readout of Y by further experimentally verifying the polarization of Y after sequential HHCP blocks and measurement of X. This can be understood by first observing in \hyperref[fig:3]{Fig.~3(a)} that we recover the initial state of X after applying a HHCP block between X and Y with spin-lock duration $L = 1/d^\text{X,Y}$, hence creating round-trip polarization transfer across the chain. This sequence is equivalent to implementing two consecutive $i$SWAP gates between X and Y. By selectively driving Y with a resonant m.w.~pulse at $\omega_0^\text{Y}$ in between the two $i$SWAP gates, the measured state of X after the second $i$SWAP will be different from the initialized state of X. Therefore any change in the signal after mapping the state of X back to the NV via HHCP(X,NV) will correspond to a single-qubit rotation of Y. This control sequence is outlined in \hyperref[fig:1]{Fig.~1(c)}, and the experimental result is shown in \hyperref[fig:3]{Fig.~3(b)}. We sweep the m.w.~pulse length to perform a Rabi experiment on Y and observe two full oscillations in the NV signal corresponding to a 4$\pi$ rotation on Y with negligible decay. This signal contrast is further evidence that Y is indeed polarized by applying two sequential HHCP blocks (NV to X, and X to Y), and the resonant m.w.~driving on Y confirms single-qubit control. Finally, the observed NV signal can be used to identify the different spin states of Y (shown via the spin-chain illustrations in \hyperref[fig:3]{Fig.~3(b)}), thus highlighting that we achieve a mechanism for readout through the NV fluorescence. With all three criteria satisfied for universal control of the second-layer spin Y, we enable solid-state registers to include previously inaccessible electronic spins in the environment of the NV center. And by implementing single-qubit quantum sensing protocols with a second-layer spin, we can increase the detection volume for single spin magnetometry through reporter spins~\cite{Sushkov2014,Schaffry2011} (see~\hyperref[sec:appendixE]{App.~E}). 
 
Finally, we characterize the optical stability of Y, as done similarly for X in a previous work~\cite{Cooper2019}. We do so by measuring the depolarization time of Y during laser illumination. \hyperref[fig:3]{Figure~3(c)} shows the polarization of Y decays with a mean lifetime of $T_{1}^\mathrm{laser} = $ 120(20) $\mu$s, enabling repetitive readout over this timescale. We can therefore harness this spin as a quantum memory and achieve simultaneous polarization of multiple environmental spins to realize more powerful sensing protocols with the NV center \cite{Jiang2009,Neumann2010}. To gain further insight into the optical properties for both the (potentially novel) X and Y defects it is necessary to perform spectroscopic measurements such as those used in~\cite{Rose2018}.

\section{EXTENDING THE SIZE OF ELECTRONIC-SPIN REGISTERS} \label{sec:scaling}

While we have restricted our spin system to include a chain up to the second layer, our results pave a clear path to extending the network to spins in higher layers as well as to multiple spins per layer. We can apply SEDOR to probe for spins in the $N^\text{th}$ layer by concatenating ($N-1$) HHCP blocks starting with the NV up to the probing spin in the $(N^\text{th}-1)$ layer. In this way, the number of control blocks scale linearly with the number of layers. Our protocol can hence be applied recursively such that the experimental resources required to control spins in higher layers (defined by the number of simultaneous spin transitions driven) remain fixed. 

We can quantify the highest layer to which our network can be scaled to by considering the decoherence effects and control errors during round-trip polarization transfer along a chain up to a spin in the $N^\text{th}$ layer. The remaining coherence of the central spin after a series of $2N$ HHCP blocks can be estimated as follows, assuming a typical $i$SWAP gate time of $t_{\text{gate}} \sim 10 \; \mu$s, spin-lock decoherence time of $T_{1,\rho} \sim 100 \; \mu$s~\cite{Cooper2019}, and $i$SWAP fidelity $\eta$: 
\begin{equation}
\langle \sigma_x^\text{NV} \rangle \approx \eta^{2N}\exp\left(-2 t_{\text{gate}} \times N \times {T_{1,\rho}^{-1}}\right).
\end{equation}
Setting the threshold for a minimum detectable signal to 10\% of the maximum NV contrast, we find we can control a chain up to the 11$^\text{th}$ layer with perfect control ($\eta = 1$).

To consider control imperfections, we estimate the $i$SWAP fidelity by performing the calibration experiment introduced in~\hyperref[sec:appendixC]{App.~C} at an optimized driving strength during spin-locking. From this data we extract a fidelity of $\eta = 0.86$, which reduces the size of the network to 4 layers. We believe the dominant source of control error stems from off-resonant driving during spin-locking. This may arise from instabilities in the setup environment and the limited resolution for the dark spin's resonance frequency (due to power broadening in the SEDOR-ESR measurement resulting in a linewidth that can be several times larger than the coupling strength).

Crucially, our method of concatenating HHCP blocks to detect a spin in a distant layer is more robust to decoherence effects compared to concatenating SEDOR blocks, as used in other experiments~\cite{vandeStolpe2023}. This is because each control step limits the number of coherent spins to just two spins, which in addition undergo relaxation in the rotating frame with a timescale of $T_{1,\rho}$ (which is in general greater than the dephasing $T_{2}$). For concatenated SEDOR, the spins in underlying layers remain in a coherent state, leading to exponentially faster decoherence over the total sequence length, as detailed in \hyperref[sec:appendixF]{App.~F}.

The electronic spin network can be even further extended using our method by searching for additional resonant spins within a given layer by increasing the detection frequency range in the SEDOR-ESR measurement. For example, in our system we could detect two additional distinct spin defects in the first layer, labeled as X$_2$ and X$_3$ (see extended data in~\hyperref[sec:appendixG]{App.~G}). Although in total this five-spin network is spectrally distinct, our method for scaling to an N-spin network is not limited to finding N non-degenerate spins. First, selective control of a few degenerate spins within the same layer can be achieved by exploiting their generally different coupling strengths to the control spin in a previous layer. Further, since our method requires maintaining coherence only between two spins, other degenerate spins in underlying layers will be unaffected by control pulses since they can be stored in a mixed state. We also remark that there exist control solutions for degenerate spins connected in a chain and other graph structures, which require only the end spin be acted upon, relaxing requirements for wider scaling of the network~\cite{Burgarth2009,Burgarth2010}. 

\section{CONCLUSIONS} \label{sec:conclusions}

In this work we develop and experimentally demonstrate control protocols to extend the electronic-spin register beyond the coherence limit of the central NV. Our experimental demonstration shows the first step to accessing increasingly larger networks of quantum systems in otherwise incoherent ensembles of electronic spins in solids. Our method is significant as building a larger network of electronic spins is critical to realizing more powerful quantum applications. In quantum sensing, larger spin networks would help increase the metrological gain achieved using entangled states of electronic spins~\cite{Cooper2019}. In quantum communication, larger dark spin chains that are stable under optical illumination would enable efficient transfer of information for larger distributed networks and distant registers~\cite{Munro2015,Awschalom2018, Mehring2006}. Furthermore, the presented approach is complementary to existing approaches to characterizing and constructing nuclear spin registers~\cite{Bradley2019, Taminiau2014, vandeStolpe2023}. These two approaches can be combined to take advantage of their complementary strengths to enable more powerful hybrid registers, where each electronic spin in the network can access its own network of nuclear spins.

By using our method to detect single spins over increasingly larger sensing volumes in parallel with ongoing efforts to discover new defects in diamond~\cite{Peaker2018,Atumi2013,Iakoubovskii2002,Cann2009,Zhou1996}, we offer a scalable approach to controlling a practical N-qubit register. This is underscored by our experimental results finding up to five possibly distinct defects in the environment of a single NV center. We also highlight that ion implantation of various elements into solids has led to the discovery of a variety of new color centers~\cite{Iwasaki2017,Czelej2018}. We can use our double resonance protocol over deeper layers in implanted samples to identify dark paramagnetic defects that would otherwise be undetected by optical or bulk EPR methods requiring large concentrations of spins. In this way we present a clear path to expanding the number of defects which can be harnessed for electronic-spin registers for quantum information processing. 
Future work will focus on deploying these methods to access larger electron-nuclear spin networks, expanding control to the nuclear-spin degree of freedom, and creating and characterizing genuine multi-partite entangled states for environment-assisted magnetometry~\cite{Cooper2019, Sun2020}.

\vskip 0.2 in 

\textit{Author’s note: during a revision of this manuscript, a
related preprint appeared mapping a 50-nuclear spin network using concatenated double-resonance sequences up to the fifth layer~\cite{vandeStolpe2023}}.

\phantomsection
\addcontentsline{toc}{section}{ACKNOWLEDGEMENTS}

\section*{ACKNOWLEDGEMENTS} \label{sec:acknowledgements}
The authors thank Santiago Hernández-Gómez and Andrew Stasiuk for their experimental support and helpful discussions. This material is based upon work in part supported by the National Science Foundation under Grant No. PHY1734011 and Graduate Research Fellowship Program under Grant No. 4000181759. The work of A.C. was supported by the Canada First Research Excellence Fund (CFREF). 

\phantomsection
\addcontentsline{toc}{section}{APPENDIX}

\section*{APPENDIX A: ANALYTICAL SEDOR SIGNAL} \label{sec:appendixA}
Here we calculate the evolution of a two-spin system (one probing spin, $i$, and one target spin, $j$) under the SEDOR pulse sequence which consists of: (1) a $\pi/2$-pulse along $y$ on $i$ (2) evolution under the mutual dipolar interaction for a time $T/2$ (3) simultaneous $\pi$-pulses on $i$ and $j$ along $x$ (4) additional dipolar evolution for a time $T/2$ (5) a $\pi/2$-pulse along $-y$ on $i$. The spin Hamiltonian in the doubly-rotating frame takes the form in the two-spin Cartesian basis~\cite{Schweiger2001}~\footnote{all constants are in units of angular frequency, and $\hbar=1$}:
\setcounter{equation}{0}
\begin{equation}
    H_{0} = \frac{1}{2} \left(\Delta \omega^i \sigma_z^{i} + \Delta \omega^j \sigma_z^{j} + \omega_d  \sigma_z^{i} \sigma_z^{j} \right).
\end{equation}
Here the detuning is $\Delta \omega = \omega_0 - \omega_{\mathrm{mw}} $ and the dipolar coupling strength is $\omega_d = 2 \pi d$. The control Hamiltonian for either spin during the pulses takes the form (ignoring cross-talk terms): 
\begin{equation}
    H_{c,(x,y)} = \frac{1}{2} \left(\Omega_0 \sigma_{(x,y)} + \Delta \omega  \sigma_z\right),
\end{equation}
where we neglect the dipolar coupling term, $\omega_d$, since it is negligible compared to $\Delta \omega$ and $\Omega_0$. Initially, the probing spin is polarized along $z$ and the target spin is in a mixed state, which is expressed by the following density matrix (taking $I_N$ to represent the $NXN$ identity operator):
 \begin{equation}
   \rho_0 = \frac{1}{2}\left(I_2 + \sigma_z^i\right) \otimes \frac{I_2}{2}.
\end{equation}
Evolution of a state $\rho$ under any two-spin Pauli operator $A$ (with $A^2 = I_4$) for a phase $\theta$ is expressed as: 
 \begin{equation}
   \rho \xrightarrow{\theta A} e^{-i \frac{\theta}{2} A}\rho e^{i \frac{\theta}{2} A} =  \cos(\theta) \rho -i \sin(\theta)\left[A,\rho \right].
\end{equation}
where we have made use of the the Baker-Hausdoff formula in the last step~\cite{Schweiger2001}. If A is a sum of operators, the last step can be applied sequentially so long as these terms commute with each other.
\vskip 0.1 in
First, we consider the case where $\Delta \omega^i = \Delta \omega^j = 0$ (on-resonance driving for both spins). This corresponds to the SEDOR-Ramsey($i$, $j$) sequence to measure the coupling strength, $d$, between spins. The total evolution of $\sigma_z^i$ under the SEDOR sequence is then calculated below: 

\begin{align}
\begin{split}
    \sigma_z^i   &\xrightarrow{\pi/2  \sigma_y^i}  \sigma_x^i  \\
      &\xrightarrow{\frac{\omega_dT}{4}\sigma_z^{i} \sigma_z^{j} }  \cos(\omega_dT/2)\sigma_x^i + \sin(\omega_dT/2)\sigma_y^{i} \sigma_z^{j} \\
    &\xrightarrow{\pi  \sigma_x^i}  \cos(\omega_dT/2)\sigma_x^i - \sin(\omega_dT/2)\sigma_y^{i} \sigma_z^{j} \\
    &\xrightarrow{\pi  \sigma_x^j}  \cos(\omega_dT/2)\sigma_x^i + \sin(\omega_dT/2)\sigma_y^{i} \sigma_z^{j} \\
      &\xrightarrow{\frac{\omega_dT}{4}\sigma_z^{i} \sigma_z^{j} }  \cos^2(\omega_dT/2)\sigma_x^i \\
      &+2\sin(\omega_dT/2)\cos(\omega_dT/2)\sigma_y^{i} \sigma_z^{j} -  \sin^2(\omega_dT/2)\sigma_x^i.
\end{split}
\end{align}
And the density matrix becomes:
\begin{align}
\begin{split}
    &\rho_f = \frac{I_4}{4} + \frac{1}{4}\cos^2(\omega_dT/2)\sigma_x^i \\
    &+\frac{1}{2}\sin(\omega_dT/2)\cos(\omega_dT/2)\sigma_y^{i} \sigma_z^{j} - 
    \frac{1}{4}\sin^2(\omega_dT/2)\sigma_x^i .
\end{split}
\end{align}
The measured signal, $\langle \sigma_z^i \rangle$, is then:
\begin{equation}
    \langle \sigma_z^i \rangle = \mathrm{Tr}(\sigma_z^i \rho_f) =  \cos^2(\omega_dT/2) - \sin^2(\omega_dT/2) = \cos(\omega_dT).
\end{equation}
Now in order to model the SEDOR-ESR($i$, $j$) signal, consider the case when we are on resonance for the probe spin ($\Delta \omega^i = 0$), and off resonance for the target spin ($\Delta \omega^j \ne 0$) as we sweep the frequency of the recoupling $\pi$-pulse. The calculation above is unaffected until the step where the density matrix evolves under $\pi \sigma_x^j$. This will modify the $\sigma_y^{i} \sigma_z^{j}$ term in the density matrix as follows: 
\begin{align}
\begin{split}
    \sigma_y^{i} \sigma_z^{j} &\rightarrow e^{\left(-i \frac{\pi}{\Omega_0^j} H_{c,x}^j\right)} \sigma_y^{i} \sigma_z^{j} e^{\left(i \frac{\pi}{\Omega_0^j} H_{c,x}^j\right)} \\
    &= A\sigma_y^{i} \sigma_z^{j} + B \sigma_y^{i} \sigma_y^{j} + C\sigma_y^{i} \sigma_x^{j}.
\end{split}
\end{align}
The only term that will contribute to the final signal when taking the trace as above will be $ A\sigma_y^{i} \sigma_z^{j} $ (since $\left[\sigma_z^{i} \sigma_z^{j}, \sigma_y^{i} \sigma_z^{j}\right] = -i\sigma_x^i$). A is computed to be:
\begin{equation}
  A = \frac{\left(\Delta \omega^j\right)^2+\Omega_0 ^2\cos \left(\frac{\pi  \sqrt{\left(\Delta \omega^j\right)^2+\Omega_0 ^2}}{\Omega_0 }\right)}{\left(\Delta \omega^j\right)^2+\Omega_0 ^2},
\end{equation}
and the final signal is now modified as: 
\begin{align}
\begin{split}
    \langle \sigma_z^i \rangle &= \text{Tr}(\sigma_z^i \rho_f) =  \cos^2(\omega_dT/2) + A \sin^2(\omega_dT/2) = \cos^2(\omega_dT/2) \\
    &+ \left(\frac{\left(\Delta \omega^j\right)^2+\Omega_0 ^2\cos \left(\frac{\pi  \sqrt{\left(\Delta \omega^j\right)^2+\Omega_0 ^2}}{\Omega_0 }\right)}{\left(\Delta \omega^j\right)^2+\Omega_0 ^2}\right) \sin^2(\omega_dT/2).
\end{split}
\end{align}
Maximum contrast occurs when $T = \pi/\omega_d = 1/(2d)$.  

\section*{APPENDIX B: DATA ANALYSIS} \label{sec:appendixB}

We fit the individual SEDOR-ESR peaks in \hyperref[fig:2]{Fig.~2(b),(e)} to a Lorentzian function with four free parameters:
\begin{equation}
   f(x) =  b_{0} + A_{0} \cdot \frac{(\frac{\gamma}{2})^2}{(x - \omega_0)^2 +(\frac{\gamma}{2})^2 }  ,
\end{equation}
from which we find $\omega_{0}^{\mathrm{X^\downarrow}}$ = $2\pi \cdot 47.0(2)$ MHz, $\omega_{0}^{\mathrm{X^\uparrow}}$ = $2\pi \cdot 73.5(2)$ MHz, $\omega_{0}^{\mathrm{Y^\downarrow}}$ =$2\pi \cdot 44.0(2)$ MHz, and $\omega_{0}^{\mathrm{Y^\uparrow}}$ = $2\pi \cdot 77.5(2)$ MHz. The error is taken as $\gamma/2$ (half of the FWHM). 

\vskip 0.2 in
We fit the SEDOR-Ramsey signals in \hyperref[fig:2]{Fig.~2(c),(f)} to the following decaying cosine function:
\begin{equation}
   f(t) =  \frac 1 2 \left( 1 + \cos(2 \pi d_0 t) \right) e^{-t/\tau_0},
\end{equation}
where for SEDOR-Ramsey(NV,X) we fix $d_0$ to the dominant frequency in its spectrum, and for SEDOR-Ramsey(X,Y) both $d_0$ and $\tau_0$ are free parameters. 

\vskip 0.2 in
The spin-echo signal in \hyperref[fig:2]{Fig.~2(c)} is fit to the following exponential decay function with three free parameters: 
\begin{equation}\label{eqn:decay}
   f(t) =  b_{0} + A_{0} \cdot e^{-t/T_2}.
\end{equation}

The Fast Fourier transform (FFT) for the SEDOR-Ramsey signals in \hyperref[fig:2]{Fig.~2(d),(g)}, as well as for the HHCP(X,Y) signal in \hyperref[fig:3]{Fig.~3(a)} are computed using the periodogram method to return the normalized power spectral density. The dominant frequency in the spectra are found by fitting the dominant peak to a Lorentzian function of the form: 
\begin{equation}
   f(x) =  \frac{(\Delta d)^2}{(x - d_0)^2 +(\Delta d)^2 },
\end{equation}
and the uncertainty reported for $d^\text{NV,X}$, $d^\text{X,Y}$, and $d^\text{HH}$ is $\Delta d$.

\vskip 0.2 in
The fit for the HHCP(X,Y) signal in \hyperref[fig:3]{Fig.~3(a)}, and the Rabi(Y) signal in \hyperref[fig:3]{Fig.~3(b)} is a cosine function with three free parameters: 
\begin{equation}
  f(t) =  b_0 +A_0 \cdot \cos(2 \pi d_0 t),
\end{equation}
and the baseline subtraction to the HHCP(X,~Y) data is equal to $b_0+A_0 -1$ such that $\langle \sigma_z^{\text{X}}(L = 0 \; \mu s) \rangle^{\mathrm{FIT}} =1$. 
\vskip 0.2 in
For the polarization decay curve in \hyperref[fig:3]{Fig.~3(c)} we fit according to \hyperref[eqn:decay]{Eq.~(13)} with $b_0$ fixed to equal zero. The fits in \hyperref[fig:3]{Fig.~3(b),(c)} exclude points less than 300 ns due to a hardware issue for sequences of variable time.

\section*{APPENDIX C:  STATE PREPARATION AND MEASUREMENT CALIBRATION (SPAM) FOR X} \label{sec:appendixC}

\setcounter{figure}{0}
\makeatletter 
\renewcommand{\thefigure}{c\@arabic\c@figure}
\makeatother


For the experiments using X as the probing spin, we convert the directly measured NV spin polarization to the X spin polarization by renormalizing to subtract off effects of imperfect control during HHCP. We create a mapping from the NV spin contrast, $\langle \sigma_z^{\text{NV}} \rangle$, to the dark spin contrast, $\langle \sigma_z^{\text{X}} \rangle$, by using the calibration signal displayed in \hyperref[fig:c1]{Fig.~c1}. This signal measures the full 2$\pi$ evolution of the X spin between two HHCP $i$SWAP gates (where the spin lock duration $T=1/(2d)$) by sweeping the phase of the second $\pi/2$ pulse on X during the first $i$SWAP gate. We convert the corresponding NV signal from $\langle \sigma_z^{\text{NV}} \rangle$ to $\langle \sigma_z^{\text{X}} \rangle$ according to the following fit, which is a cosine function with three free parameters: 
\begin{equation}\label{eq:cosFit}
     f(x) =  b_0 +A_0\cos(2 \pi f_0 x) ,
\end{equation}
where $b_0 = $ .016(2) and $A_0 = -0.35(3)$.  We then construct the mapping as: $\langle \sigma_z^X \rangle = \frac{\langle \sigma_z^{NV} \rangle -b_0}{A_0}$.

\begin{figure}[ht!]
\includegraphics[width = \columnwidth]{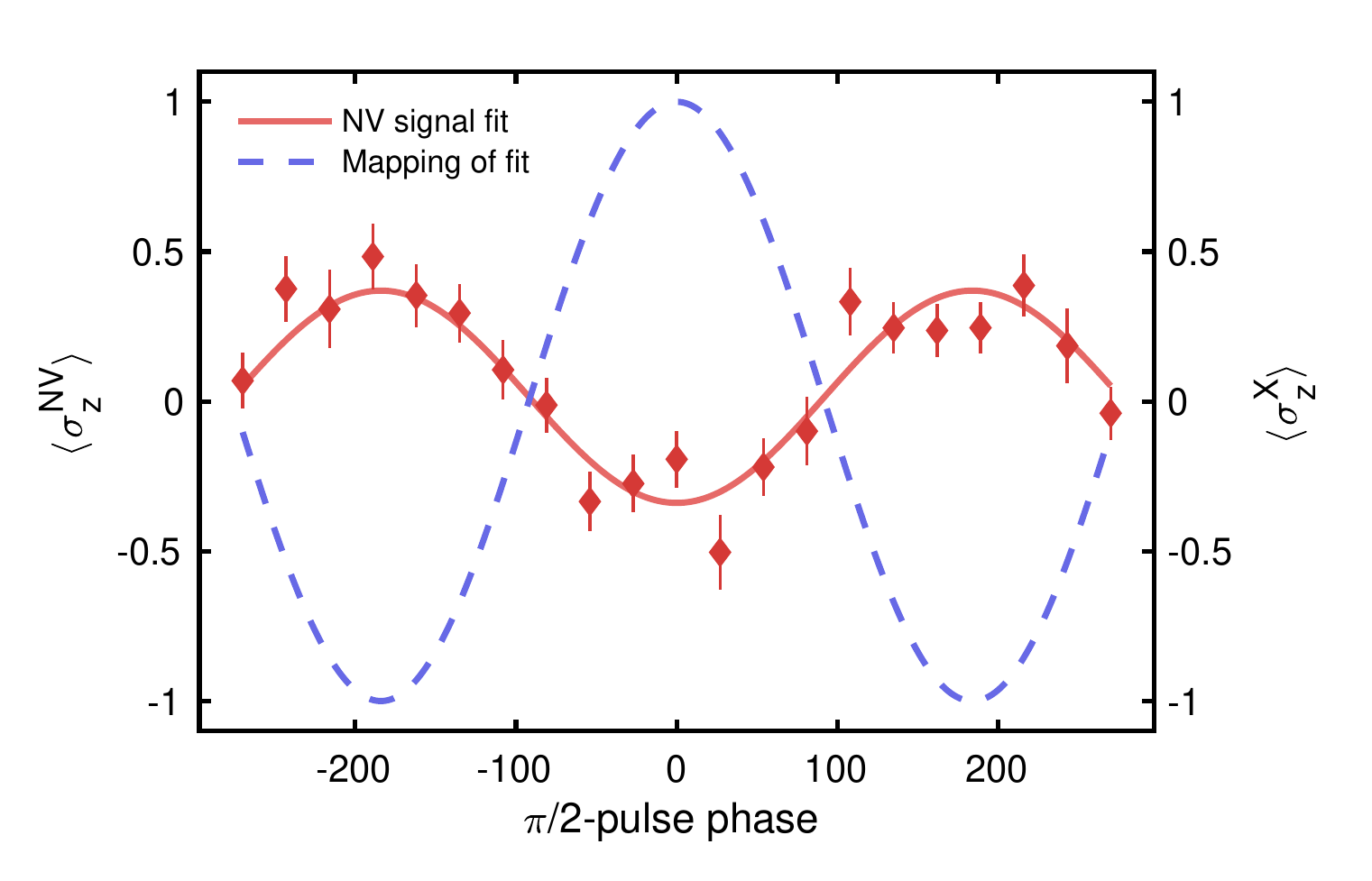}
\caption{\label{fig:c1} 
$\langle \sigma_z^{\text{NV}} \rangle$ mapping to $\langle \sigma_z^{\text{X}} \rangle$ by calibrating control errors during state preparation and measurement. Above we apply HHCP initialization and readout blocks on X, and sweep the phase of the second $\pi/2$-pulse during the initialization to evolve the X  spin state over a range of $3 \pi$ radians. Both X hyperfine transitions are driven.
}
\end{figure}

In order to include control imperfections into our estimate on network depth presented in \hyperref[sec:scaling]{Sec.~IV}, we apply the same calibration experiment but use an optimized driving strength during Hartmann-Hahn of $\Omega_0^{HH} = 1$ MHz. The resulting signal is shown in \hyperref[fig:c2]{Fig.~c2}. The round-trip SWAP efficiency is the best-fit parameter to the amplitude using the cosine fit from \hyperref[eq:cosFit]{Eq.~(16)}, which is equal to 0.74(5). This resuts in an $i$SWAP gate fidelity of $\eta = \sqrt{0.74} = 0.86$. 

\begin{figure}[ht!]
\includegraphics[width = \columnwidth]{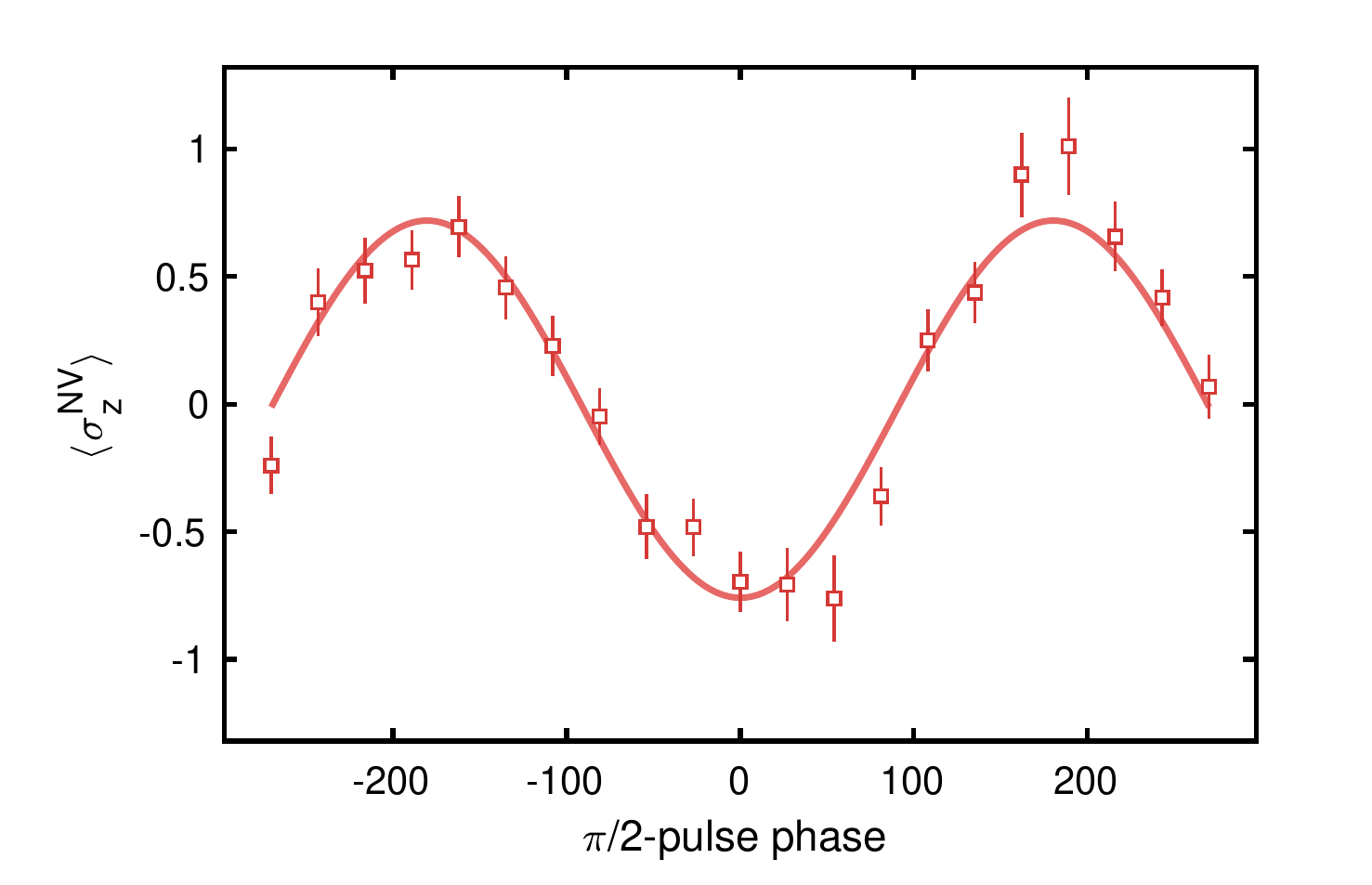}
\caption{\label{fig:c2} 
Repeating the same calibration experiment shown in \hyperref[fig:c1]{Fig.~c1}, but using twice the driving strength to achieve an optimal round-trip SWAP efficiency of 74\%.
}
\end{figure}

\section*{APPENDIX D: HYPERFINE SPLITTING OF X}\label{sec:appendixD}

Here we show that the possible hyperfine splitting values for X at any external field orientation (and thus defect orientation in the lattice) are incompatible with the measured hyperfine splitting of Y, thus concluding X and Y have different defect structures. The hyperfine contribution in the general spin Hamiltonian is:
\begin{equation}
    H_{\text{hyperfine}} = \vec{S}^{T}\cdot A \cdot \vec{I},
\end{equation}
where $\vec{S}, \vec{I}$ are the electronic and nuclear spin operators, respectively, and $A$ is the hyperfine tensor. For a uniaxially symmetric hyperfine interaction (which is the case for the NV and both X spins~\cite{Cooper2020}), $A$ can be expressed by its principal components as:
\begin{equation}
    A = \begin{pmatrix}
A_\perp &  & \\
 & A_\perp &  \\
& & A_\parallel \\
\end{pmatrix}.
\end{equation}
The hyperfine splitting, $A_s$, measured in the SEDOR-ESR spectrum can be calculated as~\cite{WeilBolton2007}:
\begin{equation}
    A_s = \sqrt{\vec{n}^{T} \cdot A \cdot A^T \cdot \vec{n}},
\end{equation}
where $\vec{n}$ defines the orientation of the external magnetic field relative to the principal direction of the hyperfine matrix. This can be expressed as a function of the external field orientation (or polar angle $\theta$ with respect to the principal axes frame), which is~\cite{Cann2009}:
\begin{equation}
    A_{s}(\theta) = \sqrt{A_{\perp}^2 \sin(\theta)^2 + A_{\parallel}^2\cos(\theta)^2}.
\end{equation}
The hyperfine components for X (X$_1$) are reported in~\cite{Cooper2020}, with $A_{\perp} = 17.2(3)$ MHz and $A_{\parallel} = 29.4(2)$ MHz. The maximum hyperfine splitting occurs at $\theta = 0$ when the external magnetic field is aligned along the principal hyperfine direction, and corresponds to $A_{s}^\text{max} = A_{\parallel}$. We assume Y is also axially symmetric, and we measure a hyperfine splitting equal to 33.5(3) MHz, which is greater than the maximum hyperfine splitting of X. Thus, we conclude X and Y must have different hyperfine tensors and therefore correspond to different defect structures. 

\section*{APPENDIX E: DETECTION VOLUME FOR SENSING WITH SPIN-CHAINS} \label{sec:appendixE}
As discussed in the main text, addressing spins in higher levels increases the effective detection volume of the (central) NV spin. 
Here we  estimate  the total detection volume for a chain of spins. The total detection volume for a chain is the sum of the individual spins’ coherence volumes minus their overlap. The volume upper bound is achieved when the detection volume of the n-th layer spin does not overlap with that of the (n-2)th layer and the spins are at the edge of each other's detection volume.  The coherence volume for a probing spin is determined by the minimum coupling strength to a target spin, set by the probing spin's coherence time through $T_{2}d(r)\sim 1$. The magnetic dipole interaction strength  $d(r)$ scales with distance as $d(r)\sim (\mu_{0}/4\pi)\gamma_e^2 \hbar r^{-3}$.
From these two relations we find the coherence volume of a single spin: $V_{0} = (4\pi/3)r^3 \sim (1/3)\mu_{0}\gamma_e^2 \hbar T_{2}$. Assuming each spin in the chain has approximately the same coherence time given their similar environments, then the total detection volume is approximately $N(2/3)V_{0}$ (the total volume for 3 spheres linked at their centers is close to twice the volume of an individual sphere).
Concretely, 
assuming a coherence time of 50 $\mu$s for each spin, the coherence volume for a single spin is a sphere with a radius of $\sim23$~nm. Then the detection range along the direction of the chain increases from $\sim46$~nm to $\sim69$~nm by extending control from the first layer to the second layer.

\section*{APPENDIX F: DECOHERENCE EFFECTS FOR SCALING TO HIGHER LAYERS} \label{sec:appendixF}
Our method relies on HHCP blocks to not only detect but also polarize higher-layer spins for their further exploitation as qubits and sensors. HHCP provides an additional benefit in terms scalability, as we discuss now by comparing the decoherence scaling with layer number, $N$, of our method of concatenating HHCP blocks  with a multi-spin correlation scheme, such as concatenating SEDOR blocks. For each control step for the concatenated HHCP method, the decoherence effects are limited to the two spins involved during each $i$SWAP gate with relaxation times $T_{1,\rho}$ and $T_{1}$. We estimate the remaining NV coherence after a sequence of round-trip $i$SWAP gates (of length $t_{\text{gate}}$) up to a spin in the $N$-th layer by:
\begin{multline}
    \langle \sigma_x^\text{NV} \rangle \approx \exp\{-2 t_{\text{gate}} \sum_{i=1}^{N}\left( T_{1,\rho}^{-1} + T_{1}^{-1}\right)\}. \\
   = \exp\{-2 t_{\text{gate}} \times N  \left( T_{1,\rho}^{-1} + T_{1}^{-1}\right)\}.
\end{multline}

In contrast, concatenating SEDOR sequences (also with length $t_{\text{gate}}$) will lead to $T_{2}$ decay for each spin pair while additionally building up a larger multi-spin coherent state subject to $T_{1}$ relaxation for all correlated qubits. The remaining NV coherence will then be:
\begin{multline}
   \langle \sigma_x^\text{NV} \rangle \approx \exp\{-2 t_{\text{gate}} \sum_{i=1}^{N}\left( T_{2}^{-1} + iT_{1}^{-1}\right)\}. \\
   = \exp\{-2 t_{\text{gate}} \times N \left( T_{2}^{-1} + (N+1)T_{1}^{-1}/2\right)\}.
\end{multline}
Thus, by direct comparison with the concatenated SEDOR approach, our approach scales exponentially better with spin-chain length with respect to $T_1$ in addition to the coherence improvement from $T_{1,\rho}>T_{2}$.

\section*{APPENDIX G: IDENTIFICATION OF TWO ADDITIONAL FIRST-LAYER SPINS X$_2$ and X$_3$} \label{sec:appendixG}
Here we show extended data of SEDOR-ESR performed on the NV over a different detection frequency window near the free electron larmor frequency, $\gamma_eB$. We identify the resonance frequencies for two additional defects in the first layer, called X$_2$ and X$_3$ (X$_1$ corresponds to X in the main text). Like X$_1$, we conclude they are both S = 1/2, I = 1/2 defects based on the contrast for each peak. The measurements were taken at a different field strength of 353 G, compared to 363 G for the experiments in the main text.

\setcounter{figure}{0}
\makeatletter 
\renewcommand{\thefigure}{g\@arabic\c@figure}
\makeatother

\begin{figure}[ht!]
\includegraphics[width = 1.1\columnwidth]{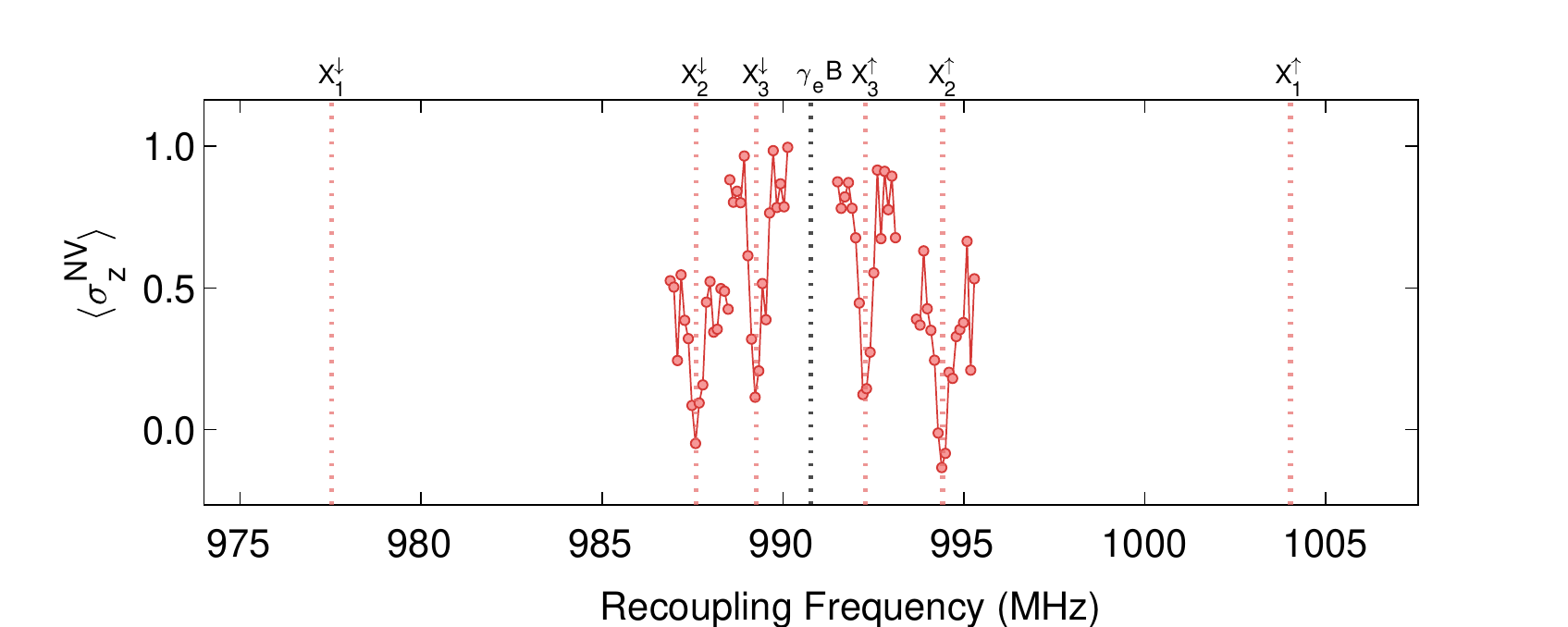}
\caption{\label{fig:g3} 
Performing the SEDOR-ESR(NV,X) experiment (introduced in~\hyperref[sec:systemA]{Sec.~II.A} of the main text) showing two additional spin defects in the first layer, labeled as X$_2$ and X$_3$. The expected X$_1$ resonance frequencies at this field strength are included as a reference.
}
\end{figure}

\bibliography{references}



\end{document}